\documentclass[aps, prl, twocolumn, showpacs, superscriptaddress, floatfix]{revtex4}

\usepackage{graphicx}
\usepackage{dcolumn}
\usepackage{bm}
\usepackage{times}

\begin{document}

\title{Exotic magnetism on the quasi-FCC lattices of the $d^3$ double perovskites La$_2$NaB$'$O$_6$ (B$'$~$=$~Ru, Os)}

\author{A.A. Aczel}
\altaffiliation{author to whom correspondences should be addressed: E-mail:[aczelaa@ornl.gov]}
\affiliation{Quantum Condensed Matter Division, Neutron Sciences Directorate, Oak Ridge National Laboratory, Oak Ridge, TN 37831, USA}
\author{P.J. Baker}
\affiliation{ISIS Facility, STFC Rutherford Appleton Laboratory, Harwell Oxford, Oxfordshire, OX11 0QX, UK }
\author{D.E. Bugaris}
\affiliation{Department of Chemistry and Biochemistry, University of South Carolina, Columbia, SC 29208, USA}
\author{J. Yeon}
\affiliation{Department of Chemistry and Biochemistry, University of South Carolina, Columbia, SC 29208, USA}
\author{H.-C. zur Loye}
\affiliation{Department of Chemistry and Biochemistry, University of South Carolina, Columbia, SC 29208, USA}
\author{T. Guidi}
\affiliation{ISIS Facility, STFC Rutherford Appleton Laboratory, Harwell Oxford, Oxfordshire, OX11 0QX, UK }
\author{D.T. Adroja}
\affiliation{ISIS Facility, STFC Rutherford Appleton Laboratory, Harwell Oxford, Oxfordshire, OX11 0QX, UK }
\affiliation{Physics Department, University of Johannesburg, P.O. Box 524, Auckland Park 2006, South Africa}

\date{\today}

\begin{abstract}
We find evidence for long-range and short-range ($\zeta$~$=$~70~\AA~at 4~K) incommensurate magnetic order on the quasi-face-centered-cubic (FCC) lattices of the monoclinic double perovskites La$_2$NaRuO$_6$ and La$_2$NaOsO$_6$ respectively. Incommensurate magnetic order on the FCC lattice has not been predicted by mean field theory, but may arise via a delicate balance of inequivalent nearest neighbour and next nearest neighbour exchange interactions. In the Ru system with long-range order, inelastic neutron scattering also reveals a spin gap $\Delta$~$\sim$~2.75~meV. Magnetic anisotropy is generally minimized in the more familiar octahedrally-coordinated $3d^3$ systems, so the large gap observed for La$_2$NaRuO$_6$ may result from the significantly enhanced value of spin-orbit coupling in this $4d^3$ material.  
\end{abstract}


\maketitle

\renewcommand{\topfraction}{0.85}
\renewcommand{\textfraction}{0.05}
\renewcommand{\floatpagefraction}{0.9}

There has been a plethora of recent interest in studying B-site ordered double perovskites (DP) with the formula A$_2$BB$'$O$_6$. When magnetic atoms only occupy the B$'$ sites and nearest neighbour (NN) antiferromagnetic (AFM) coupling is dominant, geometric frustration on the face-centered cubic (FCC) lattice is realized. Since the B and B$'$ sites can accommodate a wide variety of magnetic ions, double perovskites are particularly attractive for systematic magnetic studies of frustrated FCC systems where one can tune either the $d$ electron configuration or the spin-orbit coupling (SOC). 

A wide range of magnetic ground states have been predicted theoretically in $4d$ and $5d$ DPs with $d^1$ and $d^2$ electronic configurations\cite{10_chen, 11_chen} by considering the combined effects of strong electron correlations and strong SOC. Several exotic magnetic ground states have also been observed experimentally, including a collective singlet state/valence bond glass in Ba$_2$YMoO$_6$\cite{10_aharen, 11_carlo, 10_devries}, spin freezing without long-range order in Ba$_2$YReO$_6$\cite{10_aharen_2}, Sr$_2$MgReO$_6$\cite{03_wiebe} and Sr$_2$CaReO$_6$\cite{02_wiebe}, a ferromagnetic (FM) Mott insulating state in Ba$_2$NaOsO$_6$\cite{02_stitzer, 07_erickson}, and the $J_{eff}$~$=$~1/2 Mott insulating state in the iridates La$_2$MgIrO$_6$ and La$_2$ZnIrO$_6$\cite{13_cao}. Quantum fluctuations presumably play a large role in determining the magnetic ground state of these systems\cite{10_chen}, especially when NN AFM exchange is strong and the systems are highly-frustrated. 

In the context of the interplay between geometric frustration and SOC, there has been less interest in $4d$ and $5d$ DPs with the electronic configuration $d^3$. One downside is that $d^3$ systems are generally assumed to possess spin-only $S$~$=$~3/2 ground states with quenched orbital angular momentum according to the usual $L-S$ coupling scheme, since the magnetic B$'$ ions are in a local octahedral environment, and this configuration should minimize the effects of SOC. 

Another issue is $d^3$ DP systems are expected to behave more classically due to the large spins, and for almost all known cases long-range magnetic order is found\cite{11_greedan}. Although magnetic order cannot be stabilized on the FCC lattice solely by NN AFM exchange interactions $J_1$~$>$~0, finite next nearest neighour (NNN) exchange $J_2$ or magnetic anisotropy can alleviate the classical ground state degeneracy\cite{03_kuzmin} and allow the systems to order. The phase diagram of the $J_1$-$J_2$ model has been determined theoretically for the FCC lattice using mean field theory (MFT)\cite{62_terharr, 01_lefmann}. Four different collinear magnetic phases are found depending on the sign and magnitude of $J_1$ and $J_2$, including ferromagnetism and Type I, Type II, and Type III antiferromagnetism. All four phases have been realized in $d^3$ and $d^5$ DPs, with Type I and Type II AFM representing the most common scenarios (e.g. see Refs.~\cite{83_battle, 84_battle, 89_battle, 03_battle, 06_martin, 13_aczel_2, 02_munoz, 04_bos}). On the other hand, Type III AFM and ferromagnetic (FM) order are rather uncommon, but they have been found in the systems Ba$_2$LaRuO$_6$\cite{83_battle} and Ca$_2$SbCrO$_6$\cite{07_retuerto} respectively. 
    
Recently, we investigated the magnetism of the monoclinic $d^3$ DPs La$_2$NaRuO$_6$ and La$_2$NaOsO$_6$ by magnetic susceptibility, heat capacity and neutron powder diffraction (NPD)\cite{13_aczel}. The magnetic susceptibility shows a deviation from the Curie-Weiss law ($\theta_{CW}$~$=$~-57~K) at a temperature of 15 K for the Ru system, accompanied by a $\lambda$ anomaly in the specific heat at the same temperature. While the magnetic susceptibility of the Os system shows a similar deviation from Curie-Weiss law behaviour ($\theta_{CW}$~$=$~-74~K) around 12 K, only a broad feature is observed in the specific heat. Furthermore, in contrast to the expected collinear magnetic ground states for $d^3$ systems, we found incommensurate long-range order in La$_2$NaRuO$_6$ with a moment size of 1.87~$\mu_B$ and no magnetic Bragg peaks for La$_2$NaOsO$_6$ down to 4~K\cite{13_aczel}. This behaviour is difficult to understand in the general context of $d^3$ DPs.

In this letter, we have investigated these $d^3$ systems with muon spin relaxation ($\mu$SR) and time-of-flight neutron scattering measurements. $\mu$SR allows for a careful study of the $T$-dependence of the magnetism in these materials, while neutron scattering is useful for understanding detailed information on the nature of the magnetic ground states and spin dynamics. Our study confirms incommensurate long-range magnetic order in La$_2$NaRuO$_6$ with $T_N$~$=$~15(1)~K and reveals short-range incommensurate order in La$_2$NaOsO$_6$ down to 4~K with a correlation length $\zeta$~$=$~70~\AA. These two systems have large monoclinic $\beta$ angles relative to most other B-site ordered, $d^3$ DPs\cite{13_aczel}. While the local cubic symmetry of the magnetic B$'$ ions stays nearly intact and the B$'$ sublattice remains close to ideal FCC in monoclinic systems, the resulting structural distortions can induce substantial tilting of the BO$_6$ and B$'$O$_6$ octahedra, leading to significantly altered NN B$'$-O-O-B$'$ and NNN B$'$-O-B-O-B$'$ extended superexchange interactions. This effect seems to push La$_2$NaRuO$_6$ and La$_2$NaOsO$_6$ to the MF phase boundary between Type I and Type III AFM.  

For the La$_2$NaRuO$_6$ system with long-range magnetic order, we find a sizable spin gap $\Delta$~$\sim$~2.75~meV in the excitation spectrum. Recent neutron work has also found spin gaps in several other ordered $4d^3$ and $5d^3$ cubic and monoclinic DPs. We find that the gaps roughly scale with $T_N$, suggesting a common origin. Any plausible explanation should be based on the intermediate-to-large SOC expected in these systems. The two most likely scenarios are related to symmetric exchange anisotropy or the breakdown of $L-S$ coupling in these $4d^3$ and $5d^3$ materials. The latter could lead to a significantly unquenched orbital moment.
 
To perform the present study, polycrystalline La$_2$NaRuO$_6$ and La$_2$NaOsO$_6$ were prepared according to the procedure in Refs.~\cite{04_gemmill, 05_gemmill, 13_aczel}. For the $\mu$SR experiments, the samples were mounted onto a silver plate using GE-varnish. We used the EMU spectrometer in longitudinal geometry at the ISIS Pulsed Neutron and Muon Source, UK. The time-evolution of the spin polarization of the muon ensemble was measured via the asymmetry function, $A(t)$\cite{muSR}. The neutron scattering measurements were carried out on the MERLIN\cite{MERLIN_technical} and LET\cite{LET_technical} time-of-flight chopper spectrometers at the ISIS facility. The powder samples (mass~$\sim$~15 g) were wrapped in thin Al foil and then mounted inside a thin-walled Al can. Data was collected at temperatures between 4 and 105 K with various selected neutron energies $E_i$ between 10 and 20 meV. 

\begin{figure}
\centering
\scalebox{0.52}{\includegraphics{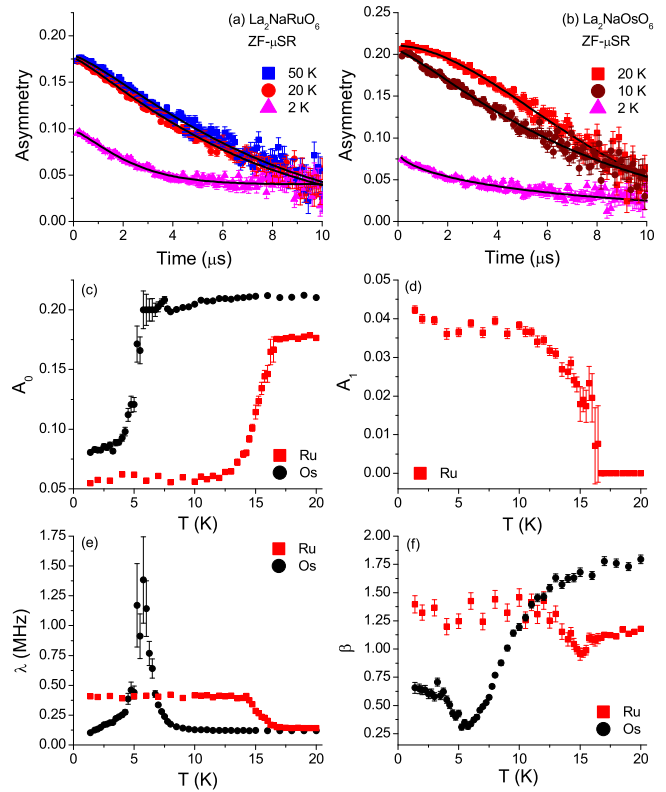}}
\caption{\label{Fig1} ZF-$\mu$SR measurements of the DPs La$_2$NaRuO$_6$ and La$_2$NaOsO$_6$. (a), (b) Asymmetry vs. time at selected temperatures for the Ru and Os systems respectively. (c) $T$-dependence of the asymmetry $A_0$ for both systems. (d) $T$-dependence of $A_1$ for the Ru system. (e), (f) $T$-dependence of the relaxation rate $\lambda$ and power $\beta$ for both systems.}
\end{figure} 

Typical zero-field (ZF) $\mu$SR data for La$_2$NaRuO$_6$ and La$_2$NaOsO$_6$ are shown in Fig.~\ref{Fig1}(a) and (b). There is a clear drop in the initial asymmetry on cooling below $T_N$~$=$~15~K and 6~K for the Ru and Os systems respectively, as shown in Fig.~\ref{Fig1}(c). This implies that internal fields larger than those that can be resolved at ISIS ($\sim$~80~mT) are present in both samples at low temperature. To model the $T$-dependence of the muon data we used the function: $A(t) = A_0 e^{-(\lambda t)^\beta} + A_1 $, where $A_0$ represents the amplitude of the relaxing signal, $\lambda$ is the muon spin relaxation rate, $\beta$ reflects the type of field distribution in the sample, and $A_1$ is a non-relaxing component only required for the Ru case at low $T$. 

For the Ru system, the drop in initial asymmetry is accompanied by the development of a non-relaxing $A_1$ component, as illustrated in Fig~\ref{Fig1}(d). In the ordered state, the muons that experience a quasi-static local field along their spin direction do not precess and therefore give rise to this non-relaxing component.  

The decrease in the initial asymmetry is also coincident with an increase in $\lambda$ and an abrupt change in $\beta$, as displayed in Fig.~\ref{Fig1}(e) and (f). The increase in $\lambda$ with decreasing $T$ is likely caused by the broad distribution of static local fields found in the long-range incommensurate state. Above $T_N$ the muon spin relaxation is exponential, as the Ru spin fluctuations enter the motional narrowing regime but stay within the frequency range for which muons remain sensitive.   

For the Os system, as shown in Fig.~\ref{Fig1}(e) the ZF-relaxation rate peaks at 6~K. This does not correspond to the 12~K ordering temperature inferred from the magnetic susceptibility and specific heat\cite{13_aczel}. It is however apparent in Fig.~\ref{Fig1}(f) that $\beta$ begins to decrease significantly below 12~K, dropping to around 1/3 at 6~K before recovering to $\sim$~2/3 at 1.5~K. This is not the usual behaviour for a system entering a long-range ordered state, and instead suggests that the Os spins are slowing down gradually and freezing below $T_f$~$=$~6~K. At higher temperatures, the relaxation is close to Gaussian, indicating that the electronic spin fluctuations are very rapid and the muon is only sensitive to fields from nuclear dipole moments. 

\begin{figure}
\centering
\scalebox{0.44}{\includegraphics{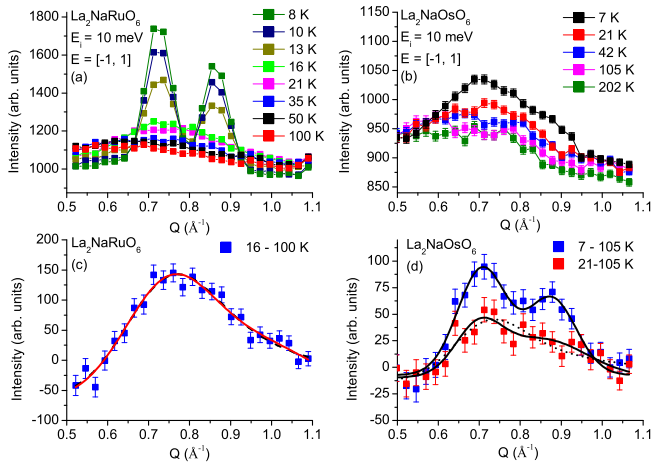}}
\caption{\label{Fig2}Elastic neutron scattering intensity for La$_2$NaRuO$_6$ and La$_2$NaOsO$_6$ from MERLIN with $E_i$~$=$~10~meV integrated over an energy range $\pm$1~meV. (a), (b) $T$-dependence of the scattering at selected temperatures for the Ru and Os systems respectively. (c) Diffuse scattering observed at 16~K for La$_2$NaRuO$_6$ with a background at 100 K subtracted. (d) Diffuse scattering at 7 and 21 K for La$_2$NaOsO$_6$ with a background at 105 K subtracted. The solid and dashed curves shown in (c) and (d) are fits described in the text.}
\end{figure} 

Time-of-flight neutron scattering provides complementary information to the $\mu$SR study. Data from MERLIN in the elastic channel with $E_i$~$=$10~meV is shown for La$_2$NaRuO$_6$ in Fig.~\ref{Fig2}(a), and reveals resolution-limited, incommensurate Bragg peaks at $Q$~$\sim$~0.72~\AA$^{-1}$ and 0.86~\AA$^{-1}$ in agreement with observations from Ref.~\cite{13_aczel}. These peaks disappear at $T_N$ and give way to diffuse scattering that decreases gradually with increasing temperature. A 16 - 100 K difference plot of the scattering is shown in Fig.~\ref{Fig2}(c). The shape of the diffuse scattering is characteristic of a Warren lineshape for 2D magnetic correlations\cite{90_zhang, 99_wills}, and short range order of this type has recently been reported above $T_N$ for the double perovskite Sr$_2$YRuO$_6$\cite{13_granado}.

The dashed black curve in Fig.~\ref{Fig2}(c) is a fit to a Warren lineshape with $Q_0$~$=$~0.73~\AA$^{-1}$~and a correlation length $\zeta$~$=$~25~\AA. The position of maximum scattering intensity $Q_0$ should correspond to the ($hk$) indices of the Bragg rod giving rise to the 2D correlations. Since the closest commensurate reflections to $Q_0$ are (0.5 0.5)$_{hl}$ and (01)$_{hl}$ with $Q$~$\sim$~0.70~\AA$^{-1}$ and 0.79~\AA$^{-1}$ respectively, a Warren lineshape does not seem to explain the diffuse scattering. Another possibility is the diffuse scattering is composed of two incommensurate magnetic peaks that are not resolution limited. To estimate the correlation length in this case, the diffuse scattering in Fig.~\ref{Fig2}(c) was fit to two Gaussian functions (solid red curve). The correlation length was then calculated using: $\zeta$~$=$~2$\pi/\sqrt{F_{HT}^2-F_{LT}^2}$, where $F_{HT}$ and $F_{LT}$ are the full-width half maximums (FWHM) of the Gaussian peaks above $T_N$ and at 8~K respectively. A value of $\zeta$~$=$~20~\AA~at 16~K is obtained by this method.    

Elastic neutron scattering results are shown in Fig.~\ref{Fig2}(b) and (d) for La$_2$NaOsO$_6$. No resolution-limited magnetic Bragg peaks are observed down to 7 K, but similar diffuse scattering is observed up to $\theta_{CW}$. The dashed black curve in (d) shows that in principle the 21-105~K Os diffuse scattering can be fit to a Warren lineshape, and in this case $\zeta$~$=$~30~\AA~and $Q_0$~$=$~0.70~\AA, corresponding to a (0.5 0.5)$_{hl}$ Bragg rod. However, this data can be fit equally well to two broad incommensurate Gaussian peaks (solid black curve) centered about the (001) Bragg position. Using the formula for $\zeta$ given above with $F_{LT}$ taken as the 8~K Ru value, this model yields $\zeta$~$=$~35~\AA. The diffuse scattering becomes two well-defined incommensurate peaks below the $T^*$~$=$~12~K bulk characterization features with $\zeta$~$=$~50~\AA~at 7 K, indicating that those anomalies correspond to a substantial increase in the correlation length of the magnetic order. 

Unlike most B-site ordered $d^3$ DPs, the magnetic ground states of La$_2$NaRuO$_6$ and La$_2$NaOsO$_6$ are not predicted by MFT. Considering the theoretical $J_1$-$J_2$ phase diagram for FCC magnets given in Ref.~\cite{01_lefmann}, one possible explanation is these systems are on the border between Type I and Type III AFM. The phase boundary corresponds to NN $J_1$~$>$~0 and a NNN $J_2$~$=$~0. This situation presumably arises due to the large tilting of the NaO$_6$ and B$'$O$_6$ octahedra weakening the ferromagnetic $J_2$ interactions necessary for Type I AFM. This scenario is more likely than the systems lying on the border between FM and Type I AFM, since they are highly-frustrated (not expected in the case of $J_1$~$=$~0 and $J_2$~$<$~0) and possess large, negative $\theta_{CW}$'s. We do not consider placing La$_2$NaRuO$_6$ and La$_2$NaOsO$_6$ on the other phase boundaries in the mean field diagram, as isostructural La$_2$LiRuO$_6$ with a smaller monoclinic distortion is Type I AFM\cite{03_battle}, and in general most Ru$^{5+}$/Os$^{5+}$ DPs are Type I AFM. 

\begin{figure}
\centering
\scalebox{0.12}{\includegraphics{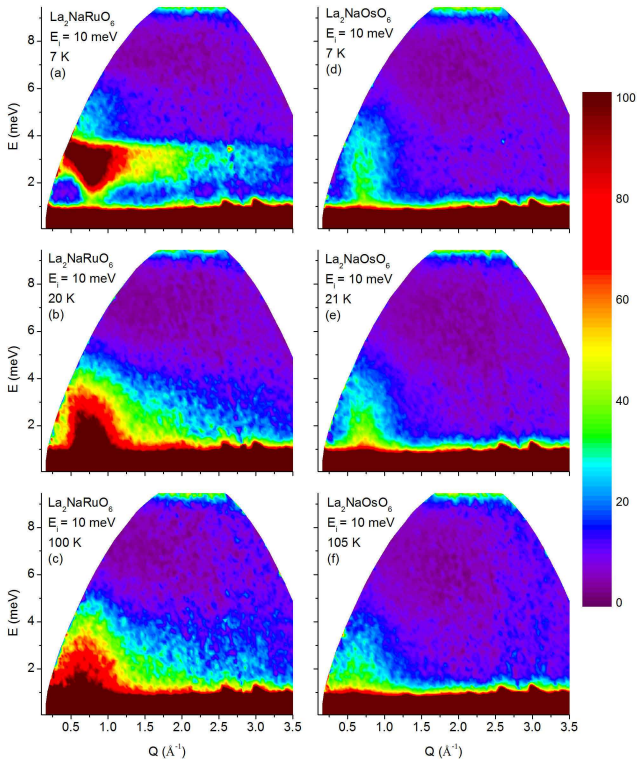}}
\caption{\label{Fig3} Neutron scattering spectra for (a)-(c) La$_2$NaRuO$_6$ and (d)-(f) La$_2$NaOsO$_6$ at selected temperatures from MERLIN with $E_i$~$=$~10~meV. As the temperature drops below $T_N$~$=$~15~K for the Ru sample, a spin gap with $\Delta$~$\sim$~2.75~meV opens. The onset of a spin gap for the Os sample appears below $T^*$~$=$~12~K, but this never becomes well-defined.}
\end{figure} 

\begin{figure}
\centering
\scalebox{0.44}{\includegraphics{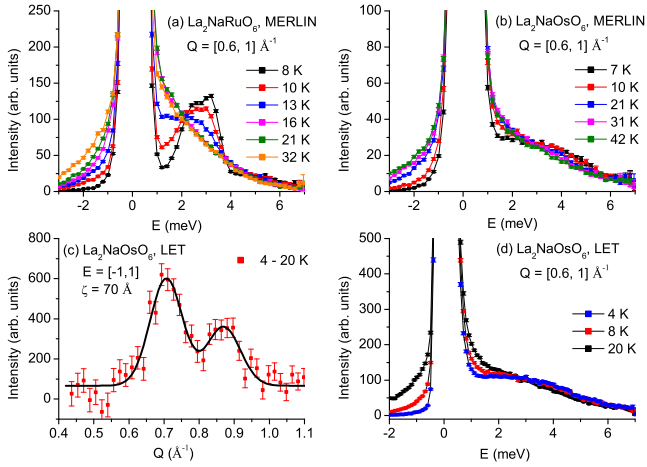}}
\caption{\label{Fig4} Scattering intensity vs. energy transfer from MERLIN with $E_i$~$=$~10~meV, integrated over 0.6~$<$~$Q$~$<$~1~\AA$^{-1}$ for (a) La$_2$NaRuO$_6$ and (b) La$_2$NaOsO$_6$. The opening of a spin gap below $T_N$ is clear for the long-range ordered Ru system. (c) Difference plot of elastic neutron scattering data for La$_2$NaOsO$_6$, integrated over -1~$<$~$E$~$<$ 1, from LET with $E_i$~$=$~10~meV. The two short-range incommensurate peaks persist down to 4~K. (d) Scattering intensity vs. energy transfer, integrated over 0.6~$<$~$Q$~$<$~1~\AA$^{-1}$, for La$_2$NaOsO$_6$ from LET with $E_i$~$=$~10~meV. A spin gap never fully opens down to 4~K, in agreement with the absence of long-range magnetic order.} 
\end{figure}

Color maps of the neutron scattering spectra are shown in Fig.~\ref{Fig3}(a)-(c) for La$_2$NaRuO$_6$ at selected temperatures. Below T$_N$, a clear spin gap of $\Delta$~$\sim$~2.75~meV opens up in the inelastic channel. A spin wave bandwidth of $\sim$~4~meV is apparent by examining a low temperature $E_i$~$=$~20~meV dataset (not shown). The $T$-evolution of the spin gap is illustrated more clearly in Fig.~\ref{Fig4}(a), which shows the energy dependence of the $Q$-integrated scattering around the incommensurate $Q$~$=$~0.72~\AA$^{-1}$ and 0.86~\AA$^{-1}$ positions (0.6 $<$ $Q$ $<$ 1~\AA$^{-1}$). The observation of a spin gap with a magnitude larger than $T_N$ in an octahedrally-coordinated $d^3$, 3D system is highly unusual. For the more familiar case of 3$d^3$ systems, a combination of strong crystal fields and negligible SOC generally ensure that the magnetic anisotropy is minimal.

Additional insight on the origin of the spin gap for La$_2$NaRuO$_6$ results by direct comparison to the spin gaps observed for other cubic and monoclinic DPs. The cubic 4$d^3$ system Ba$_2$YRuO$_6$ ($T_N$~$=$~36~K) has $\Delta$~$=$~5~meV \cite{13_carlo}, the cubic 5$d^3$ system Ba$_2$YOsO$_6$ ($T_N$~$=$~70~K) has $\Delta$~$=$~15~meV \cite{gaulin}, and the monoclinic 4$d^3$ system Sr$_2$YRuO$_6$ ($T_N$~$=$~24~K) has $\Delta$~$=$~5~meV \cite{13_adroja}. The transition temperatures and gap sizes seem to roughly scale with one another, suggesting a common origin. The cubic crystal fields for Ru$^{5+}$/Os$^{5+}$ in Ba$_2$YRuO$_6$ and Ba$_2$YOsO$_6$, combined with the quenched orbital angular momentum expected from the $L-S$ coupling scheme, rule out single ion anisotropy as as an origin of the spin gap in those cases\cite{01_alders}. The high cubic symmetry of these two compounds also eliminates the Dzyaloshinsky-Moriya (DM) interaction from consideration. 

Symmetric exchange anisotropy is a second-order SOC effect involving the excited states of two magnetic ions, and therefore usually much weaker compared to single ion anisotropy and the DM interaction. However, it can play an important role in the magnetic anisotropy of $4d^3$ and $5d^3$ DPs\cite{03_kuzmin}, where these other effects are minimized and SOC is significant. In fact, recent work on monoclinic Sr$_2$YRuO$_6$ has shown that the gapped magnetic excitation spectrum can be explained well with a model that includes NN symmetric exchange anisotropy\cite{13_adroja}. This effect is even allowed by symmetry in the case of the cubic systems with perfect magnetic FCC sublattices\cite{86_halg}, and therefore could be the primary spin gap mechanism in all these DPs.
 
On the other hand, the spin gaps may arise from the breakdown of $L-S$ coupling in these $4d$ and $5d$ systems. Recent theoretical work\cite{13_matsuura} shows that SOC values typical of $4d$ and $5d$ transition metals, combined with reduced intra-Coulomb interactions due to the extended orbitals, lead to an unquenched orbital moment and magnetic anisotropy in octahedrally-coordinated $4d^3$ and $5d^3$ systems. It is proposed that these materials belong to a regime intermediate between $L-S$ and $J-J$ coupling. 

In contrast to La$_2$NaRuO$_6$, a spin gap never fully develops below $T_f$ down to 4~K for La$_2$NaOsO$_6$, as indicated with combined MERLIN and LET data presented in Fig.~\ref{Fig3}(d)-(f) and Fig.~\ref{Fig4}(b) and (d). The MERLIN data allows for a direct comparison with the Ru sample, while the LET data explores the $T$~$<$~$T_f$ regime. The LET data in Fig.~\ref{Fig4}(c) shows that the short-range order persists down to 4~K and the incommensurate peaks become more clearly defined below $T^*$. The FWHM of the 4~K peaks corresponds to a correlation length $\zeta$ of 70~\AA. Note that $\zeta$ was calculated using the same method as for the 7 K Os data from MERLIN, with $F_{LT}$ obtained from the magnetic peaks of 4~K La$_2$NaRuO$_6$ LET data. These observations further illustrate that the gapped excitations in $4d^3$ and $5d^3$ double perovskites can be associated directly with the long-range order. The different magnetic behaviour of La$_2$NaRuO$_6$ and La$_2$NaOsO$_6$ is intriguing, and the key to a more complete understanding may lie in future theoretical investigations of extended superexchange interactions and the magnetism on the phase boundaries of the FCC MFT $J_1$-$J_2$ phase diagram. 

In conclusion, muon spin relaxation and neutron scattering measurements find evidence for long-range and short-range incommensurate magnetic order on the quasi-FCC lattices of the monoclinic DPs La$_2$NaRuO$_6$ and La$_2$NaOsO$_6$ respectively. These magnetic states may arise due to a delicate balance of exchange interactions induced by the large tilting of the NaO$_6$ and B$'$O$_6$ octahedra. Furthermore, in the Ru $d^3$ system with long-range order, inelastic neutron scattering reveals a spin gap $\Delta$~$\sim$~2.75~meV. The values of $T_N$ and the magnitude of the gaps in ordered $4d^3$ and $5d^3$ DPs seem to exhibit nearly linear scaling behaviour, suggesting a common origin. We propose that these spin gaps arise as a consequence of the intermediate-to-large SOC in these materials, through either symmetric anisotropic exchange or the breakdown of $L-S$ coupling. X-ray magnetic circular dichroism measurements on $4d^3$ and $5d^3$ cubic DPs are essential to distinguish between these two possibilities.  

\begin{acknowledgments}
We acknowledge B.D. Gaulin, J.P. Clancy and J.P. Carlo for useful discussions. This research was supported by the US Department of Energy, Office of Basic Energy Sciences. A.A.A. was supported by the Scientific User Facilities Division. D.E.B. and H.z.L. would like to acknowledge financial support through the Heterogeneous Functional Materials for Energy Systems (HeteroFoaM) Energy Frontiers Research Center (EFRC), funded by the US Department of Energy, Office of Basic Energy Sciences under award number DE-SC0001061.
\end{acknowledgments}


\begin{thebibliography}{99}
\bibitem{10_chen}G. Chen, R. Pereira and L. Balents, Phys. Rev. B {\bf 82}, 174440 (2010).
\bibitem{11_chen}G. Chen and L. Balents, Phys. Rev. B {\bf 84}, 094420 (2011).
\bibitem{10_aharen}T. Aharen, J.E. Greedan, C.A. Bridges, A.A. Aczel, J. Rodriguez, G.J. MacDougall, G.M. Luke, T. Imai, V.K. Michaelis, S. Kroeker, H.D. Zhou, C.R. Wiebe and L.M.D. Cranswick, Phys. Rev. B {\bf 81}, 224409 (2010).
\bibitem{11_carlo}J.P. Carlo, J.P. Clancy, T. Aharen, Z. Yamani, J.P.C. Ruff, J.J. Wagman, G.J. Van Gastel, H.M.L. Noad, G.E. Granroth, J.E. Greedan, H.A. Dabkowska and B.D. Gaulin, Phys. Rev. B {\bf 84}, 100404(R), (2011).
\bibitem{10_devries}M.A. de Vries, A.C. Mclaughlin and J.W.G. Bos, Phys. Rev. Lett. {\bf 104}, 177202 (2010).
\bibitem{10_aharen_2}T. Aharen, J.E. Greedan, C.A. Bridges, A.A. Aczel, J. Rodriguez, G.J. MacDougall, G.M. Luke, V.K. Michaelis, S. Kroeker, C.R. Wiebe, H.D. Zhou and L.M.D. Cranswick, Phys. Rev. B {\bf 81}, 064436 (2010).
\bibitem{03_wiebe}C.R. Wiebe, J.E. Greedan, P.P. Kyriakou, G.M. Luke, J.S. Gardner, A. Fukaya, I.M. Gat-Malureanu, P.L. Russo, A.T. Savici and Y.J. Uemura, Phys. Rev. B {\bf 68}, 134410 (2003).
\bibitem{02_wiebe}C.R. Wiebe, J.E. Greedan, G.M. Luke and J.S. Gardner, Phys. Rev. B {\bf 65}, 144413 (2002).
\bibitem{02_stitzer}K.E. Stitzer, M.D. Smith and H.-C. zur Loye, Solid State Science {\bf 4}, 311 (2002).
\bibitem{07_erickson}A.S. Erickson, S. Misra, G.J. Miller, R.R. Gupta, Z. Schlesinger, W.A. Harrison, J.M. Kim and I.R. Fisher, Phys. Rev. Lett. {\bf 99}, 016404 (2007).
\bibitem{13_cao}G. Cao, A. Subedi, S. Calder, J.-Q. Yan, J. Yi, Z. Gai, L. Poudel, D.J. Singh, M.D. Lumsden, A.D. Christianson, B.C. Sales and D. Mandrus, Phys. Rev. B {\bf 87}, 155316 (2013).
\bibitem{11_greedan}J.E. Greedan, S. Derakhshan, F. Ramezanipour, J. Siewenie and Th. Proffen, J. Phys.: Cond. Matt. {\bf 23}, 164213 (2011).
\bibitem{03_kuzmin}E.V. Kuz'min, S.G. Ovchinnikov and D.J. Singh, Phys. Rev. B {\bf 68}, 024409 (2003).
\bibitem{62_terharr}D. ter Harr and M.E. Lines, Philos. Trans. R. Soc. London A {\bf 254}, 521 (1962); {\bf 255}, 1 (1962).
\bibitem{01_lefmann}K. Lefmann and C. Rischel, Eur. Phys. J. B {\bf 21}, 313 (2001).
\bibitem{83_battle}P.D. Battle, J.B. Goodenough and R. Price, Journal of Solid State Chemistry {\bf 46}, 234 (1983).
\bibitem{84_battle}P.D. Battle and W.J. Macklin, Journal of Solid State Chemistry {\bf 52}, 138 (1984).
\bibitem{89_battle}P.D. Battle and C.W. Jones, Journal of Solid State Chemistry {\bf 78}, 108 (1989).
\bibitem{03_battle}P.D. Battle, C.P. Grey, M. Hervieu, C. Martin, C.A. Moore and Y. Paik, Journal of Solid State Chemistry {\bf 175}, 20 (2003).
\bibitem{06_martin}L. Ortega-San Martin, J.P. Chapman, L. Lezama, J.S. Marcos, J. Rodriguez-Fernandez, M.I. Arriortua and T. Rojo, Eur. J. Inorg. Chem. 1362 (2006). 
\bibitem{13_aczel_2}A.A. Aczel, D.E. Bugaris, J. Yeon, C. de la Cruz, H.-C. zur Loye and S.E. Nagler, Phys. Rev. B {\bf 88}, 014413 (2013).
\bibitem{02_munoz}A. Munoz, J.A. Alonso, M.T. Casais, M.J. Martinez-Lope and M.T. Fernandez-Diaz, J. Phys.: Cond. Matt. {\bf 14}, 8817 (2002).
\bibitem{04_bos}J.-W.G. Bos and J.P. Attfield, Phys. Rev. B {\bf 70}, 174434 (2004).
\bibitem{07_retuerto}M. Retuerto, M. Garcia-Hernandez, M.J. Martinez-Lope, M.T. Fernandez-Diaz, J.P. Attfield and J.A. Alonso, Journal of Materials Chemistry {\bf 17}, 3555 (2007).
\bibitem{13_aczel}A.A. Aczel, D.E. Bugaris, L. Li, J.-Q. Yan, C. de la Cruz, H.-C. zur Loye and S.E. Nagler, Phys. Rev. B {\bf 87}, 014435 (2013).
\bibitem{04_gemmill}W.R. Gemmill, M.D. Smith and H.-C. zur Loye, Journal of Solid State Chemistry {\bf 177}, 3560 (2004).
\bibitem{05_gemmill}W.R. Gemmill, M.D. Smith, R. Prozorov and H.-C. zur Loye, Inorg. Chem. {\bf  44}, 2639 (2005).
\bibitem{muSR}S.J. Blundell, Contemporary Physics {\bf 40}, 175 (1999).
\bibitem{MERLIN_technical}R.I. Bewley, R.S. Eccleston, K.A. McEwen, S.M. Hayden, M.T. Dove, S.M. Bennington, J.R. Treadgold and R.L.S. Coleman, Physica B {\bf 385-386}, 1029 (2006).
\bibitem{LET_technical}R.I. Bewley, J.W. Taylor and S.M. Bennington, Nucl. Instrum. Meth. A {\bf 637}, 128 (2011).
\bibitem{90_zhang}H. Zhang, J.W. Lynn, W.-H. Li and T.W. Clinton, Phys. Rev. B {\bf 41}, 11229 (1990).
\bibitem{99_wills}A.S. Wills, N.P. Raju, C. Morin and J.E. Greedan, Chem. Mater. {\bf 11}, 1936 (1999).
\bibitem{13_granado}E. Granado, J.W. Lynn, R.F. Jardim and M.S. Torikachvili, Phys. Rev. Lett. {\bf 110}, 017202 (2013).
\bibitem{13_carlo}J.P. Carlo, J.P. Clancy, K. Fritsch, C.A. Marjerrison, G.E. Granroth, J.E. Greedan, H.A. Dabkowska and B.D. Gaulin, Phys. Rev. B {\bf 88}, 024418 (2013).
\bibitem{gaulin}B.D. Gaulin {\it et al}, in preparation.
\bibitem{13_adroja}D.T. Adroja, J. Paddison, R. Singh, C.V. Tomy, M. Rotter, P. Deen, W. Kockleman, M. Koza, J.R. Stewart and A. Goodwin, in preparation.
\bibitem{01_alders}D. Alders, R. Coehoorn and W.J.M. de Jonge, Phys. Rev. B {\bf 63}, 054407 (2001).
\bibitem{86_halg}B. Halg and A. Furrer, Phys. Rev. B {\bf 34}, 6258 (1986). 
\bibitem{13_matsuura}H. Matsuura and K. Mityake, J. Phys. Soc. Jpn. {\bf 82}, 073703 (2013).
\end{thebibliography}
\end{document}